\documentclass[twocolumn,prl,showpacs]{revtex4}
	
\newcommand{\beq}{\begin{eqnarray}}
\newcommand{\eeq}{\end{eqnarray}}

\usepackage{graphicx,amssymb}

\begin{document}
\title{Occupancy of phase space, extensivity of $S_q$, and $q$-generalized  central limit theorem
}

\author{
Constantino Tsallis  \footnote{tsallis@santafe.edu}
}
\address{
Santa Fe Institute, 1399 Hyde Park Road, Santa Fe, New Mexico 87501,  USA \\
and\\
Centro Brasileiro de Pesquisas F\'\i sicas,\
Rua Xavier Sigaud 150, 
22290-180 Rio de Janeiro-RJ, Brazil
}
\date{\today}

\begin{abstract}
Increasing the number $N$ of elements of a system typically makes the entropy to increase. The question arises on {\it what particular entropic form} we have in mind and {\it how it increases}  with $N$. Thermodynamically speaking it makes sense to choose an entropy which increases {\it linearly} with $N$ for large $N$, i.e., which is {\it extensive}. If the $N$ elements are probabilistically {\it independent} (no interactions) or quasi-independent (e.g., {\it short}-range interacting), it is known that the entropy which is extensive is that of Boltzmann-Gibbs-Shannon, $S_{BG} \equiv -k \sum_{i=1}^W p_i \ln p_i$. If they are however {\it globally correlated} (e.g., through {\it long}-range interactions), the answer depends on the particular nature of the correlations. There is a large class of correlations (in one way or another related to scale-invariance) for which an appropriate entropy is that on which nonextensive statistical mechanics is based, i.e., $S_q \equiv k \frac{1-\sum_{i=1}^W p_i^q}{q-1}$ ($S_1=S_{BG}$), where $q$ is determined by the specific correlations. We briefly review and illustrate these ideas through simple examples of occupation of phase space. A very similar scenario emerges with regard to the central limit theorem (CLT). If the variables that are being summed are {\it independent} (or quasi-independent, in the sense that they gradually become independent if $N\to\infty$), two basic possibilities exist: if the variance of the random variables that are being composed is {\it finite}, the $N\to\infty$ attractor in the space of distributions is a Gaussian, whereas if it {\it diverges}, it is a L\'evy distribution. If the variables that are being summed are however {\it globally correlated}, there is no reason to expect the usual CLT's to hold. The $N\to\infty$ attractor is expected to depend on the nature of the correlations. That class of correlations (or part of it) that makes $S_q$ to be extensive for $q \ne 1$ is expected to have a  $q_e$-Gaussian as its  $N\to\infty$ attractor, where $q_e$ depends on $q$ [$q_e(q)$ such that $q_e(1)=1$], and where $q_e$-Gaussians are proportional to $[1-(1-q_e)\beta \,x^2]^{1/(1-q_e)}$ ($\beta >0$; $q_e<3$). We present some numerical indications along these lines. The full clarification of such a possible connection would have considerable interest: it would help qualifying the class of systems for which the nonextensive statistical concepts are applicable, and, concomitantly, it would enlighten the reason for which $q$-exponentials are ubiquitous in many natural and artificial systems.
\end{abstract}
\maketitle

\section{1 - On the extensivity of $S_q$}

A surface is a geometric object which has a {\it finite} measure for dimensionality $d=2$, {\it zero} measure for any $d>2$, and {\it infinite} measure for any $d<2$. A fractal has a {\it finite} measure only for $d=d_f$, where the {\it fractal dimension} $d_f$ is some real number, {\it zero} measure for $d>d_f$, and {\it infinite} measure for $d<d_f$. Generically $d_f$ is a noninteger number (an exception is constituted by the so called {\it fat fractals}, which have an integer $d_f$). For example, a triadic Cantor set constructed on a segment of length $10\,cm$  has a measure of $10^{d_f}\,cm^{d_f}$ with $d_f= \ln 2/\ln 3 =0.63...$, i.e., $4.27\,cm^{0.63}$. 

In total analogy with the above, an entropy $S$ is said {\it extensive} if $\lim_{N\to\infty} S(N)/N$ is {\it finite}. Let us focus on the basis of nonextensive statistical mechanics \cite{Tsallis88} (see \cite{review1,review2} for reviews), namely on the entropy
\begin{equation}
S_q \equiv k \frac{1-\sum_{i=1}^W p_i^q}{q-1}\;\;\;\; (\sum_{i=1}^W p_i=1; \,q \in {\cal{R}}) \,,
\end{equation} 
with $S_1=S_{BG} \equiv -k \sum_{i=1}^W p_i \ln p_i$. 
There are systems, constituted by $N$ elements, whose probabilistic sets are such that  $\lim_{N\to\infty} S_q(N)/N$ is {\it finite} for $q=q_{sen}$, {\it vanishes} for $q>q_{sen}$, and {\it diverges} for $q<q_{sen}$ ({\it sen} stands for {\it sensitivity}: see \cite{TsallisEPN,RobledoEPN} and references therein). We shall here illustrate both normal (i.e., $q_{sen}=1$) and anomalous (i.e., $q_{sen} \ne1$) systems. 

Let us consider $N$ distinguishable binary random variables whose {\it joint} probabilities are denoted by $\{p_1^A,p_2^A\}\equiv \{ p, 1-p\}$ for $N=1$, $\{p_{11}^{A+B},p_{12}^{A+B}, p_{21}^{A+B}, p_{22}^{A+B} \}$ (with $p_{11}^{A+B}+p_{12}^{A+B}+ p_{21}^{A+B}+ p_{22}^{A+B}=1$) for $N=2$, and so on. See Table I for $N=3$. For arbitrary $N$ we will have a $N$-dimensional hypercube within which we can represent the associated $W=2^N$ joint probabilities. The $N=2$ {\it marginal} probabilities are given by $\bar{p}_1^A=p_{11}^{A+B}+p_{12}^{A+B}$, $\bar{p}_2^A=p_{21}^{A+B}+p_{22}^{A+B}$, $\bar{p}_1^B=p_{11}^{A+B}+p_{21}^{A+B}$, and $\bar{p}_2^B=p_{12}^{A+B}+p_{22}^{A+B}$. Analogously we can construct the marginal probabilities associated with arbitrary $N$. 
\begin{table}[htbp]
\begin{tabular}{c||c|c||}
 $_A\setminus^B$    &  1                                                                & 2                                                    \\
[1mm] \hline\hline
1                              &  $p_{111}^{A+B+C}$                                & $p_{121}^{A+B+C}$    \\   
                                & $[p_{112}^{A+B+C}]$                               &$[p_{122}^{A+B+C}]$            \\                                               
[3mm] \hline
2                              &  $p_{211}^{A+B+C}$                                & $p_{221}^{A+B+C}$             \\
                                &  $[p_{212}^{A+B+C}]$                              & $[p_{222}^{A+B+C}]$          \\                                         
[3mm] \hline \hline
\end{tabular}
\caption{Joint probabilities associated with $N=3$, the subsystems being $A$, $B$ and $C$. The probabilities out of and within brakets respectively correspond to states 1 and 2 of subsystem $C$. It is $p_{111}^{A+B+C}+p_{112}^{A+B+C}+p_{121}^{A+B+C}+p_{122}^{A+B+C}+p_{211}^{A+B+C}+p_{212}^{A+B+C}+p_{221}^{A+B+C}+p_{222}^{A+B+C}=1$.  
}
\end{table}

Let us introduce now an important notion, namely {\it scale-invariance}, basis of techniques such as renormalization group \cite{Murray}. We shall say that our system is {\it scale-invariant} if the marginal probabilities of the $N$- system coincide with the joint probabilities of the $(N-1)$- system, $\forall N$. This strong property implies, for instance, that  $\bar{p}_1^A=p_1^A$, $\bar{p}_2^A=p_2^A$, $\bar{p}_1^B=p_1^B$, and $\bar{p}_2^B=p_2^B$. 

Let us also introduce a symmetry property, namely {\it commutativity}. A $N$-system will be said {\it commutative} if we can freely interchange its subsystems for every specific $N$-state. This property implies, for instance, that  $p_{12}^{A+B}= p_{21}^{A+B}$ for $N=2$, and $p_{112}^{A+B+C}=p_{121}^{A+B+C}=p_{211}^{A+B+C}$ and $p_{122}^{A+B+C}=p_{212}^{A+B+C}=p_{221}^{A+B+C}$ for $N=3$. It does {\it not} imply $p_{11}^{A+B}= p_{22}^{A+B}$ for $N=2$, nor $p_{111}^{A+B+C}=p_{222}^{A+B+C}$ or $p_{112}^{A+B+C}=p_{221}^{A+B+C}$ for $N=3$. If a system is commutative for all $N$, it admits a representation simpler that the $N$-dimensional hypercubes mentioned before. It admits, more precisely, a triangular representation as indicated in Table II, with the convention $(r_{20},r_{21}, r_{22}) \equiv   (p_{11}^{A+B},p_{12}^{A+B},p_{22}^{A+B})$ for $N=2$, $(r_{30},r_{31}, r_{32},r_{33}) \equiv   (p_{111}^{A+B+C},p_{121}^{A+B+C},p_{212}^{A+B+C},p_{222}^{A+B+C})$ for $N=3$, and so on for increasing $N$. Probability normalization implies of course
\begin{equation}
\sum_{n=0}^N \frac{N!}{(N-n)! \,n!} \, r_{Nn}=1 \;\;\;(\forall N) \,.
\end{equation}

\begin{figure}
\begin{center}
\includegraphics[scale=0.7]{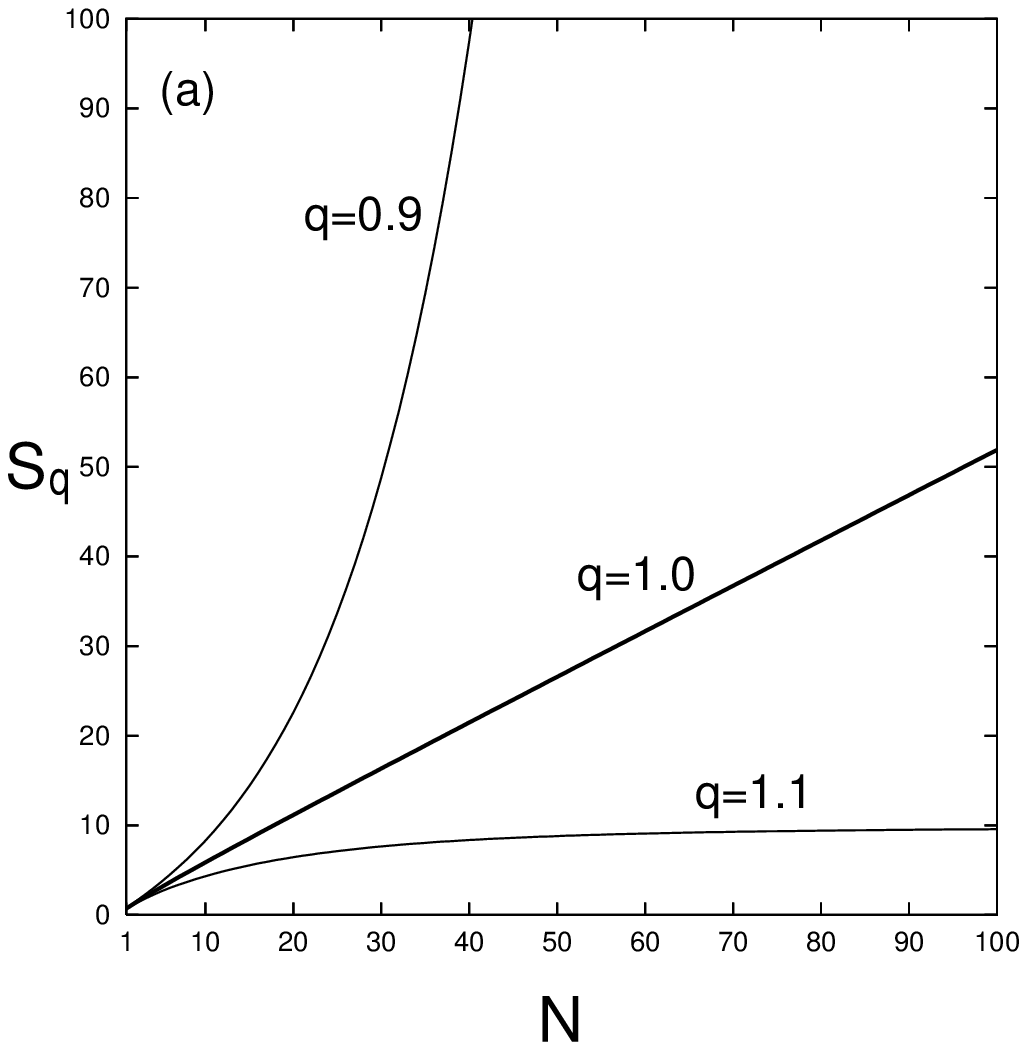} \\
\includegraphics[scale=0.7]{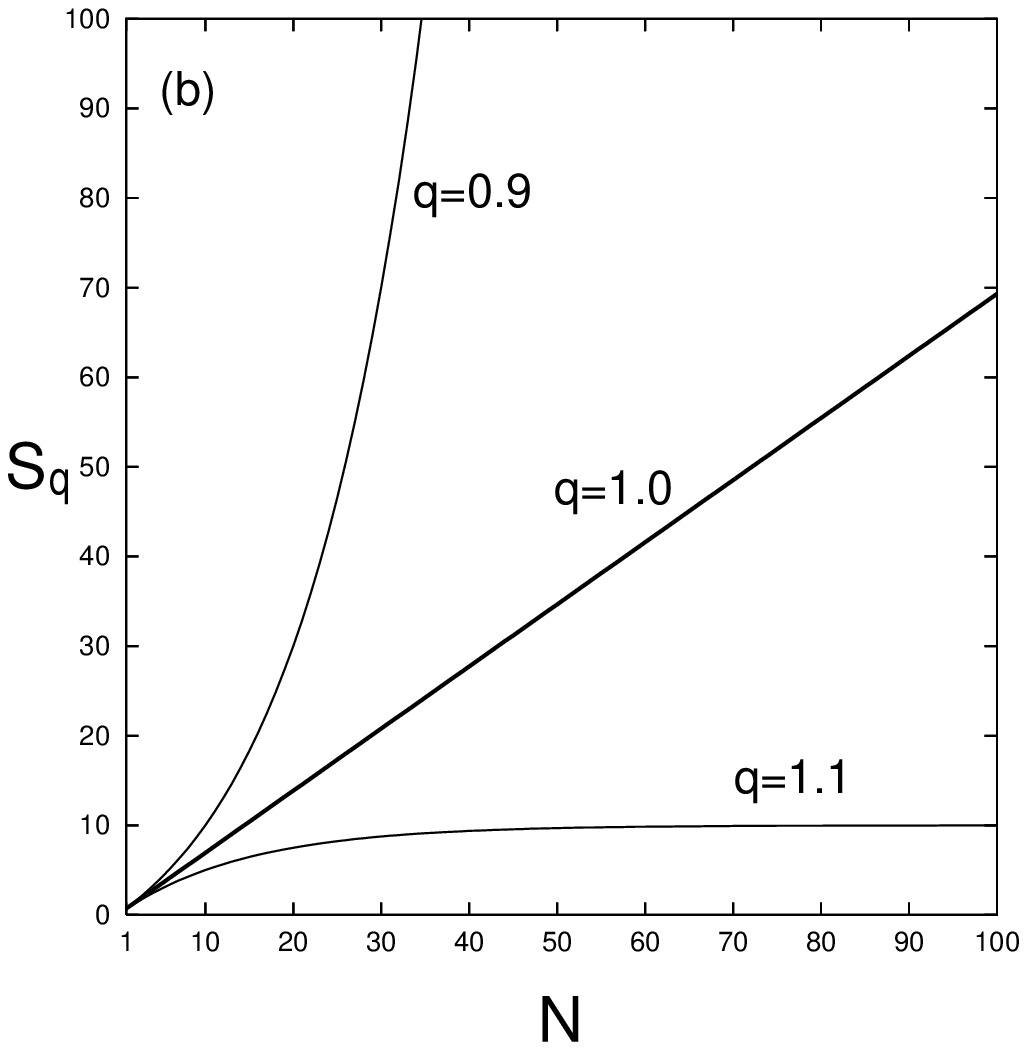} \\
\includegraphics[scale=0.7]{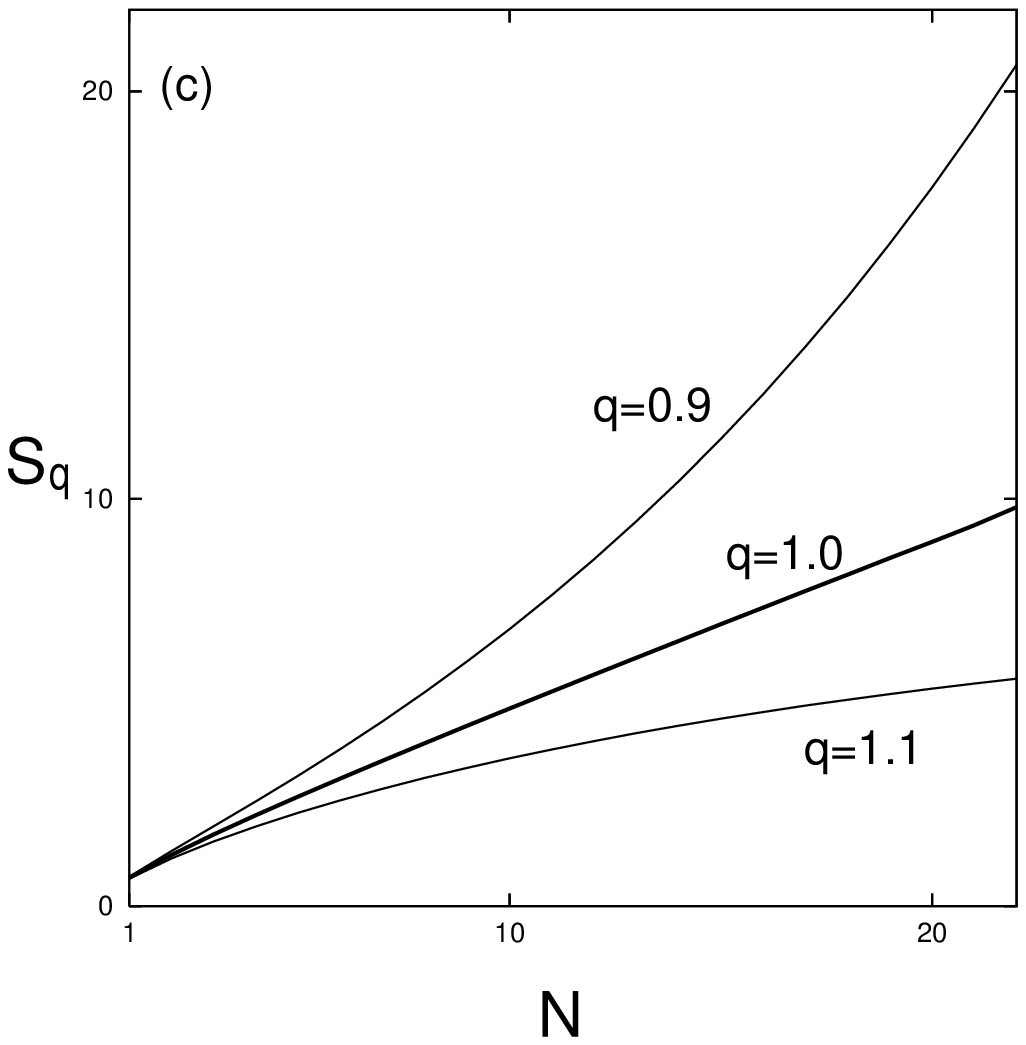} \\
\end{center}
\caption{\small $S_q(N)$ for (a) the Leibnitz triangle, (b) $p=1/2$ independent subsystems, and (c) $r_{N,0}= (1/2)^{N^{1/2}}$. Only for $q=1$ we have a {\it finite} value for $\lim_{N\to\infty}S_q(N)/N$; it {\it vanishes} ({\it diverges}) for $q>1$ ($q<1$).
}
\end{figure}

\begin{table}[htbp]
\begin{flushleft}
$(N=0)$        ~~~~~~~~~~~~~~~~~~~~~~         ~~~~$(1,1)$~~~~\\
$(N=1)$        ~~~~~~~~~~~~~~~~         ~~~$(1,r_{10})$~~$(1,r_{11})$~~~\\ 
$(N=2)$        ~~~~~~~~~~~~         ~~$(1,r_{20})$~~$(2,r_{21})$~~$(1,r_{22})$~~\\ 
$(N=3)$        ~~~~~~~         ~$(1,r_{30})$~~$(3,r_{31})$~~$(3,r_{32})$~~$(1,r_{33})$~\\ 
$(N=4)$        ~~~         $(1,r_{40})$~~$(4,r_{41})$~~$(6,r_{42})$~~$(4,r_{43})$~~$(1,r_{44})$\\
\end{flushleft}
\vspace{-0.5cm}
\caption{Most general form of a commutative system composed by binary random variables. The left (right) component of each couple corresponds to the Pascal triangle element (the probability).
}
\end{table}

Scale-invariance and commutativity are totally independent properties. If they are, however, simultaneously satisfied, then the system can be represented as in Table II with the following supplementary property:
\begin{equation}
r_{N,n}+r_{N,n+1}=r_{N-1, n} \,.
\end{equation}
This property is from now on referred to as the ``Leibnitz rule" \cite{preTGS,TGS,Marsh}. Indeed, it is satisfied by the celebrated Leibnitz triangle \cite{Polya}. The Leibnitz rule goes ``upwards", in contrast with the Pascal triangle rule, which goes ``downwards". A triangle such as that of Table II, and also satisfying Leibnitz rule, is fully determined by giving one probability for each value of $N$, e.g., by giving $\{r_{N0}\}$ ($\forall N$). Leibnitz triangle itself is entirely determined by $r_{N0}=1/(N+1)$ (hence $r_{Nn}=\frac{1}{N+1}\frac{(N-n)! \,n!}{N!}$).

Let us now give three $q_{sen}=1$ examples, all of them satisfying Leibnitz rule. They are indicated in Fig. 1, and correspond to the Leibnitz triangle itself, $N$ {\it independent} binary random variables (i.e., $r_{N0}=p^N$, hence $r_{Nn}=p^{N-n}(1-p)^n$ with $0 \le p \le 1$), and a {\it streched-exponential} system respectively. Both the Leibnitz triangle and the streched-exponential system involve correlations, but not global enough to take the system out from the $q_{sen}=1$ universality class.  In other words, the entropy which is extensive is $S_{BG}$, the Boltzmann-Gibbs one .

Let us finally illustrate the $q_{sen} \ne 1$ case with a triangle in which most states have zero probability. Only a (left-side) strip whose width is $d+1$ ($d=1,2,...$) has nonvanishing probabilities (see details in \cite{TGS}). The number of states with nonvanishing probabilities increases like a power of $N$, whereas the total number of states increases like $2^N$. The $d=1,2$ instances are presented in Table III. The entropies of the $d=1,2,3$ instances are shown in Fig. 2. The correlations are now global enough to drive the system out of the $q_{sen}=1$ universality class. The entropy which is extensive now is $S_{1-(1/d)}$. 

\begin{table}[htbp]
$(N=0)$~~~~~~~~~~~~~~~~~~~~~~~~~~$(1,1)$~~~~~~~~~~~~~~~~~~~~~~~~~~~~~~~\\
$(N=1)$~~~~~~~~~~~~~~~~~~~$(1,1/2)$~~$(1,1/2)$~~~~~~~~~~~~~~~~~~~~~~~~  \\ 
$(N=2)$~~~~~~~~~~~~~~$(1,1/2)$~~$(2,1/4)$~~$(1,0)$~~~~~~~~~~~~~~~~~~~~    \\ 
$(N=3)$~~~~~~~~~$(1,1/2)$~~$(3,1/6)$~~~$(3,0)$~~$(1,0)$~~~~~~~~~~~~~~~\\ 
$(N=4)$~~~$(1,1/2)$~~$(4,1/8)$~~~~$(6,0)$~~$(4,0)$~~$(1,0)$~~~~~~~~~~~~\\
\vspace{0.5cm}
$(N=0)$~~~~~~~~~~~~~~~~~~~~~~~~~~$(1,1)$~~~~~~~~~~~~~~~~~~~~~~~~~~~~~~~~\\
$(N=1)$~~~~~~~~~~~~~~~~~~$(1,1/2)$~~$(1,1/2)$~~~~~~~~~~~~~~~~~~~~~~~~~  \\ 
$(N=2)$~~~~~~~~~~~~$(1,1/3)$~~$(2,1/6)$~~$(1,1/3)$~~~~~~~~~~~~~~~~~~~    \\ 
$(N=3)$               ~~~~~$(1,3/8)$~$(3,5/48)$~~$(3,5/48)$~~$(1,0)$~~~~~~~~~~~~~~\\ 
$(N=4)$~                    $(1,2/5)$$(4,3/40)$$(6,3/60)$~~~~$(4,0)$~~$(1,0)$ ~~~~~~~~~\\
\caption{Anomalous probability sets: $d=1$ ({\it top}), and $d=2$ ({\it bottom}). They satisfy the Leibnitz rule only asymptotically, i.e., for $N\to\infty$.} 
\end{table}

\begin{figure}
\begin{center}
\includegraphics[scale=0.63]{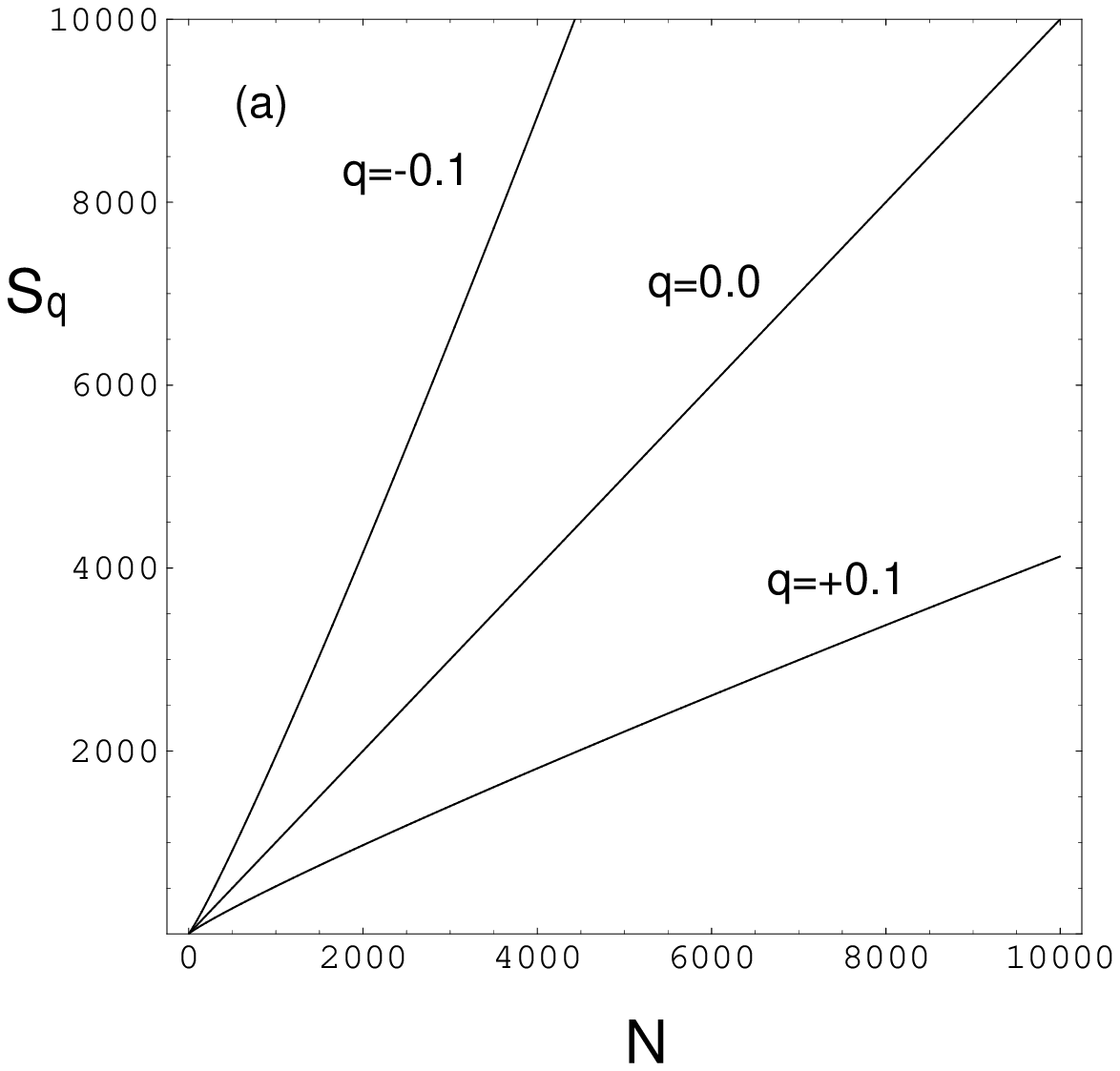} \\
\includegraphics[scale=0.63]{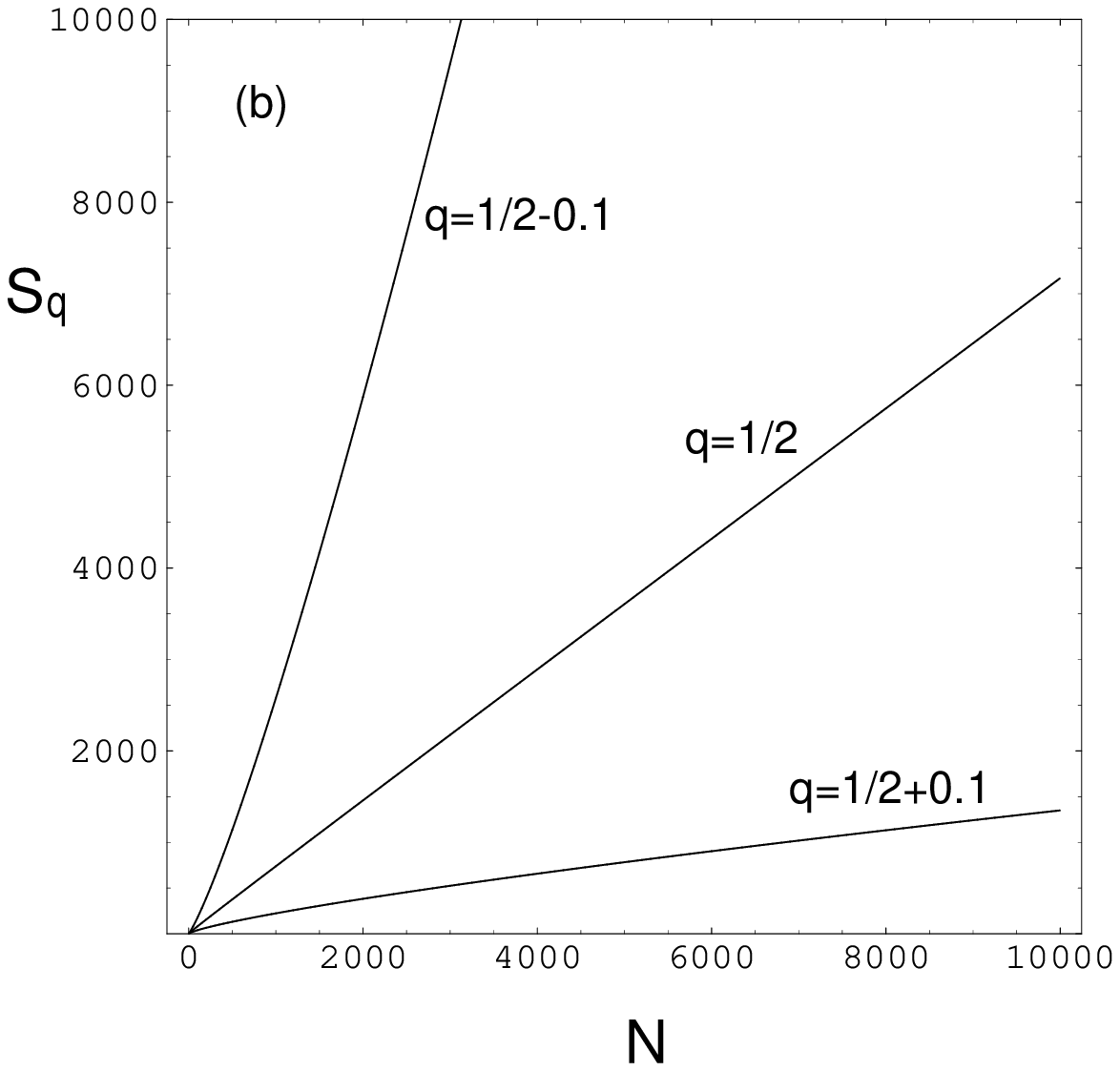} \\
\includegraphics[scale=0.63]{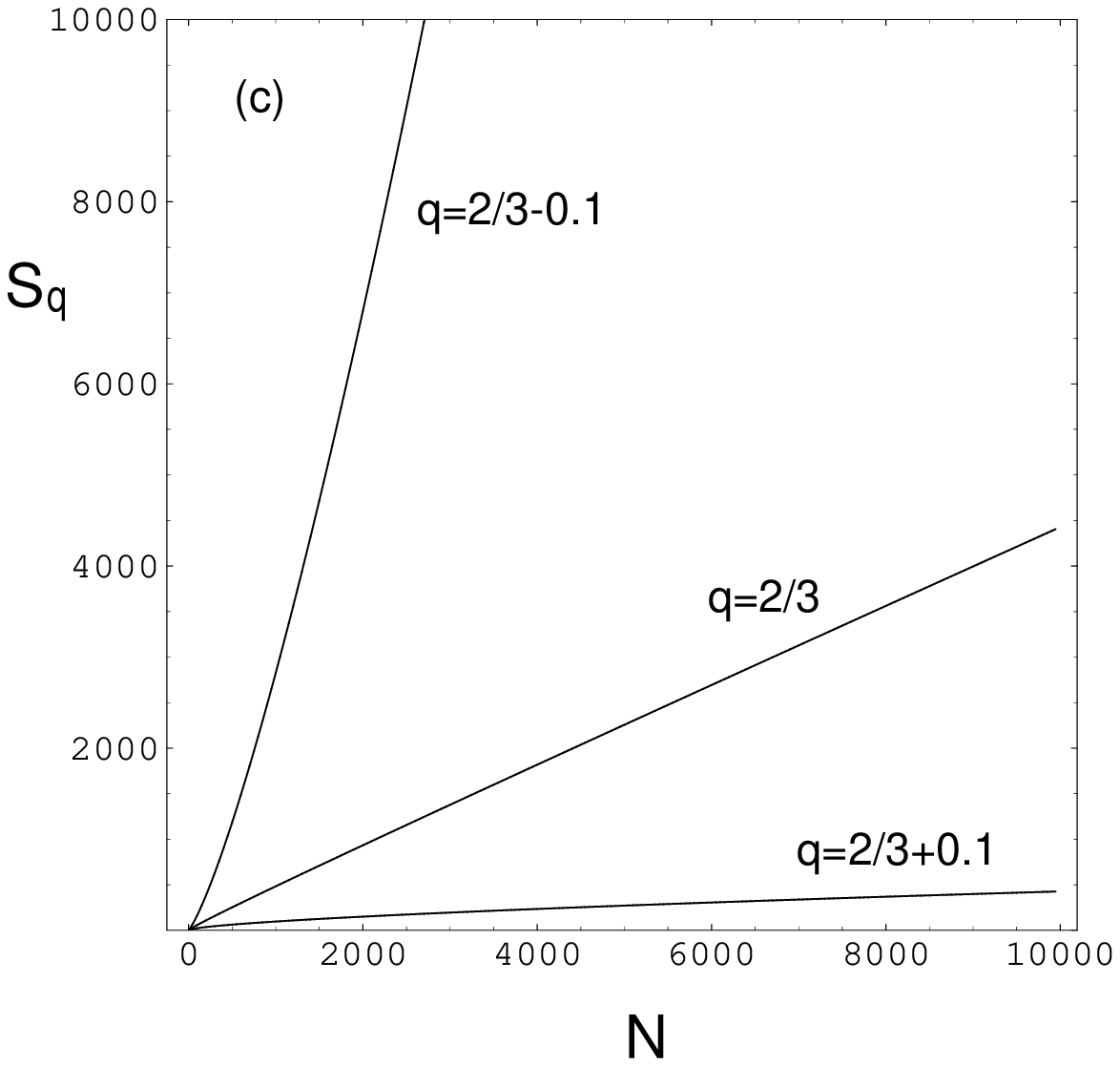} \\
\end{center}
\vspace{-0.5cm}
\caption{\small $S_q(N)$ for anomalous systems: (a) $d=1$, (b) $d=2$, and (c) $d=3$. Only for $q=1-(1/d)$ we have a {\it finite} value for $\lim_{N\to\infty}S_q(N)/N$; it {\it vanishes} ({\it diverges}) for $q>1+(1/d)$ ($q<1+(1/d)$).
}
\end{figure}

\section{2 - On a $q$-generalized Central Limit Theorem}

Let us consider the following generalized Fokker-Planck equation:
\begin{equation}
\frac{\partial p(x,t)}{\partial t}=D \frac{\partial^\gamma [p(x,t)]^{2-q_e}}{\partial |x|^\gamma} \;\;\;(0 < \gamma \le 2; \, q_e<3) \,.
\end{equation}
The diffusion coefficient $D$ can always be englobed within time $t$. Consequently, this equation is essentially characterized by only two parameters, namely $\gamma$ and $q_e$: see Fig. 3. 

If $(q_e,\gamma)=(1,2)$, we have the celebrated Fourier heat equation, and the exact solution (assuming $p(x,0)=\delta(0)$) is a Gaussian. This point of Fig. 3 is to be associated with the standard or Gaussian Central Limit Theorem ($G-CLT$). Within this theorem one essentially considers the sum of $N$ {\it independent} random variables, each of them satisfying a probability distribution whose variance is {\it finite}. Then the $N \to\infty$ attractor in the space of the distributions is, after appropriate scaling, a Gaussian distribution. $N$ plays in the theorem the same role as $t$ in Eq. (4).

If $q_e=1$ and $0< \gamma<2$, we still have a {\it linear} equation whose exact solution (assuming $p(x,0)=\delta(0)$) is a L\'evy distribution the index of which coincides with $\gamma$. This line of Fig. 3 is to be associated with the L\'evy-Gnedenko $CLT$ ($L-CLT$). Within this theorem, once again one considers the sum of $N$ {\it independent} random variables, each of them satisfying a probability distribution whose variance now {\it diverges}. Then the $N \to\infty$ attractor in the space of the distributions is, after appropriate scaling, a L\'evy distribution. 

These two theorems basically cover all possible situations whenever the random variables are {\it independent}. The situation might change dramatically if the $N$ variables are {\it collectively correlated} for all values of $N$, even when $N$ diverges. One such case is the one indicated in Fig. 3 by the $\gamma=2$ line with $q_e \ne 1$. The equation is then {\it nonlinear}, and its exact solution \cite{plastinos,bukman} (assuming $p(x,0)=\delta(0)$) is a $q_e-Gaussian$ (this name denotes the distributions that are proportional to $e_{q_e}^{-B\,x^2}$ with $B>0$, where $e_{q_e}^x \equiv [1+(1-q_e)\,x]^{1/(1-q_e)}$, and $e_1^x=e^x$); normalizability of the distribution imposes $q_e<3$; the second moment is finite for $q_e<5/3$, and diverges for $5/3 \le q_e<3$. We expect this family of solutions (quite common indeed for certain classes of complex systems) to correspond to some form of generalized $CLT$ (from now referred to as $q-CLT$). In other words, we expect that a certain class of globally correlated systems would have as their $N\to\infty$ attractor, and after some appropriate scaling,  precisely a $q_e$-Gaussian.

\begin{figure}
\begin{center}
\includegraphics[width=8.5cm,angle=0]{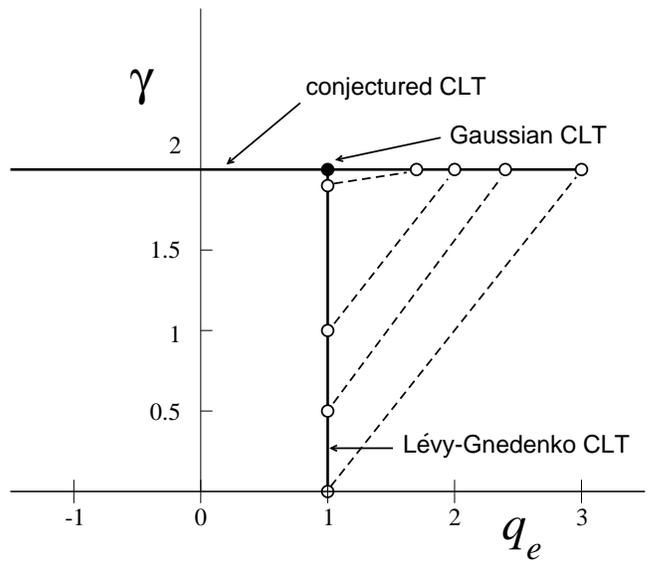}
\end{center}
\caption{\small Localization in the $(q_e,\gamma)$-space of the standard and L\'evy-Gnedenko CLT's, as well as of the conjectural $q$-generalized CLT (based on \cite{milano}. 
The schematic dashed lines are curves that share the {\it same} exponent of the power-law behavior that emerges in the limit $|x| \to \infty$. At the $q_e=1$ axis we have L\'evy distributions which asymptotically decay as $1/|x|^{1+\gamma}$, and at the $\gamma=2$ axis we have $q_e$-Gaussians which decay as $1/|x|^{2/(q_e-1)}$. The connection is therefore given by $q_e=(\gamma+3)/(\gamma+1)$ for $2>\gamma>0$, hence $5/3<q_e<3$ (see \cite{TsallisLevy}, based on \cite{ZanetteLevy}). For instance, the dashed line which joins the $(q_e, \gamma)$ points $(1,1)$ and $(2,2)$ schematically indicates those solutions of Eq. (4) which asymptotically decay as $1/x^2$, and the dashed line joining $(1,1/2)$ and $(7/3,2)$ indicates those solutions which decay as $1/|x|^{3/2}$. The dot slightly to the right of the point $(5/3,2)$ is joint to the point slightly below $(1,2)$.
}
\end{figure}

The CLT and its possible extensions have a long and fascinating history, clearly justified by the crucial role this theorem plays in theory of probabilities, statistical mechanics, and elsewhere. In some sense it all started with A. de Moivre, who in 1733 obtained (apparently for the first time \cite{stigler}) the normal distribution in connection with the binomial distribution, Pascal triangle and alike. It continued with P.S. de Laplace, who reobtained the normal distribution in 1774, and then with R. Adrain in 1808. Finally, in 1809 C.F. Gauss published his celebrated formalization of the theory of errors. It is presumably because of this achievement that it became to be known nowadays as {\it Gaussian}. In the 1930's, P. L\'evy and B.V. Gnedenko developed the extended theorem yielding L\'evy distributions. Many studies have been dedicated to variations of these theorems. Efforts have been also done addressing the influence of correlations (e.g., \cite{mello}), and reviews are available (see \cite{sornette,jonalasinio} and references therein). 

The possibility of existence of a generalized $CLT$ whose attractors would be  $q$-Gaussians was considered in \cite{bologna,tsallisbali,swinney}. Such a possibility was addressed in detail in \cite{milano}, in particular in connection with deformed products \cite{products,products2,products3,products4,products5}. The same theorem was essentially addressed in \cite{celiawalks} in terms of stochastic processes. A related theorem, also based in deformed products, was recently presented as well \cite{baldovinstella}. 

We focus now onto some recent numerical indications \cite{moyanogellmann,tsalliskyoto} suggesting the existence of the $q_e$-Gaussian attractors, at least for the $q_e \le 1$ branch of the $\gamma=2$ line of Fig. 3. More precisely, we are going to generalize the binomial distribution by introducing a global correlation (between $N$ random variables) through a specific {\it scale-invariant} procedure. We then check what is the $N \to\infty$ attractor. The absence of such correlations will provide a Gaussian attractor ($q_e=1$), which is usually referred to as the de Moivre-Laplace theorem. In other words, we intend to generalize here that theorem in the presence of a specific classs of global correlations. We follow \cite{moyanogellmann}.

We consider again the $N$ distinguishable binary variables that were introduced previously (see Table II). They can be associated  with a row of a triangle of probabilities $\{r_{N,n}\}$ ($n=0,1,2,...,N$) that satisfy Eq. (2). Once again we shall impose  {\it Leibnitz rule} \cite{preTGS,TGS}, i.e., Eq. (3).
We remind that this rule is a specific form of scale-invariance which guarantees that the {\it marginal} probabilities of the $N$- system coincide with the {\it joint} probabilities of the $(N-1)$- system. The probabilistic system is fully determined if we also provide one element of each row of the triangle, say $\{r_{N,0}\}$ ($\forall N$). If we impose $r_{N,0}=p^N$ with $0 \le p \le1$, we precisely recover the binomial distribution, which leads to the de Moivre-Laplace theorem. We shall impose a global correlation by using the $q$-product \cite{products3} defined as follows:
\begin{equation}
\label{e.qproddefinition}
x \otimes_q y \equiv [x^{1-q}+y^{1-q}-1]^{1/(1-q)} \;\;\;\;(x,y \ge 1; q \le 1).
\end{equation}
This generalised product has the following properties: (i) $x \otimes_1 y=x \, y$; (ii) $x \otimes_q 1=x $; (iii) $\ln_q (x \otimes_q y) = \ln_q x + \ln_q y$, with $\ln_q x \equiv \frac{x^{1-q}-1}{1-q} $ $(\ln_1 x=\ln x)$ being the inverse of $e_q^x$; (iv) $\frac{1}{x \otimes_q y}=(\frac{1}{x}) \otimes_{2-q}( \frac{1}{y})$. If the probability distribution in phase space is uniform within a volume $W$, the entropy $S_q$ is given by $S_q=\ln_q W$. Property (iii) can then be interpreted as $S_q(A+B)=S_q(A)+S_q(B)$ where $A$ and $B$ are subsystems that are not independent but rather satisfy $W_{A+B}=W_A \otimes_q W_B$. 
The possibility of a correspondence between this $q-product$ with a $q$-CLT has already been conjectured \cite{milano}, and some efforts along this line already exist in the literature \cite{products4}.

\begin{figure}
\begin{center}
\includegraphics[width=7.0cm,angle=-90]{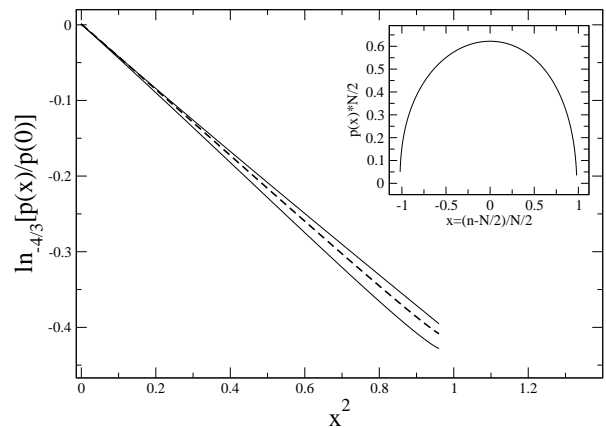}
\end{center}
\vspace{-0.5cm}
\caption{\small $ \ln_{-4/3} \frac{p(x)}{p(0)}$ {\it vs} $x^2$ for $(q,p)=(3/10,1/2)$, and $N=200$. Two branches are observed due to the asymmetry emerging from the fact that we have imposed the $q$-product on the ``left" side of the triangle; we could have done otherwise. The mean value of the two branches is indicated in dashed line. It is through this mean line that we have numerically calculated $q_e(q)$ as indicated in Fig. 5. {\it Inset:} Linear-linear representation of $p(x)$.
}
\end{figure}

\begin{figure}
\begin{center}
\includegraphics[scale=0.33]{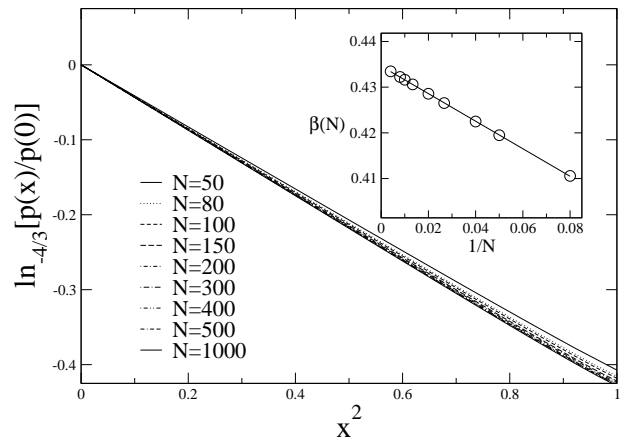} \\
\end{center}
\vspace{-0.5cm}
\caption{\small $\ln_{-4/3} \frac{p(x)}{p(0)}$ {\it vs} $x^2$ for $(q,p)=(3/10,1/2)$ and various system sizes $N$.  {\it Inset:} $N$-dependence of the (negative) slopes of the $\ln_{q_e}$ {\it vs} $x^2$ straight lines. 
We find that, for $p=1/2$ and $N>>1$, 
$\langle(n- \langle n \rangle)^2 \rangle \sim N^2/\beta(N) \sim a(q)N + b(q)N^2$. For $q=1$ we find $a(1)=1$ and $b(1)=0$, consistent with {\it normal} diffusion as expected. For $q<1$ we find $a(q)>0$ and $b(q)>0$, thus yielding {\it ballistic} diffusion. The linear correlation factor of the $q-log \,versus\, x^2$ curves range from 0.999968 up to near 0.999971 when $N$ increases from 50 to 1000. The very slight lack of linerarity that is observed is expected to vanish in the limit $N \to\infty$, but at the present stage this remains a numerically open question.
}
\end{figure}

\begin{figure}
\begin{center}
\includegraphics[scale=0.33]{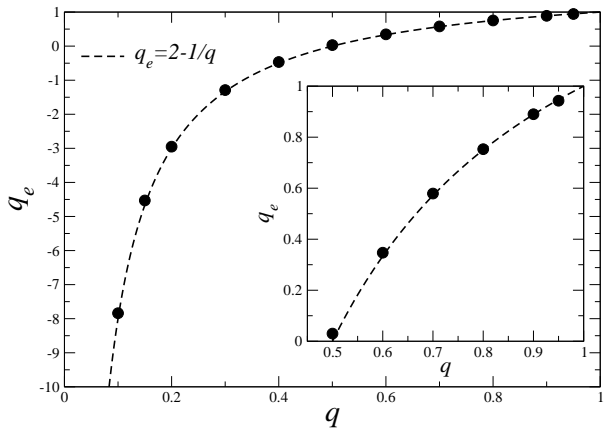} \\
\end{center}
\vspace{-0.5cm}
\caption{\small Relation between the index $q$ from the $q$-product definition, and
 the index $q_e$ resulting from the numerically calculated probability
  distribution. The agreement with the analytical conjecture $q_e=2-\frac{1}{q}$ is
  remarkable. {\it Inset:} Detail for the range $0<q_e<1$.
}
\end{figure}

The global correlation is introduced by imposing
\begin{equation}
\label{e.qproduct}
(1/r_{N,0})=(1/p)\otimes_{q} (1/p)\otimes_{q} (1/p)\otimes_{q} \ldots \otimes_{q} (1/p) \,,
\end{equation}
hence
\begin{eqnarray}
 r_{N,0}&=&p\otimes_{2-q} p\otimes_{2-q} p\otimes_{2-q} \ldots \otimes_{2-q} p \nonumber \\
&=&1/\, [Np^{\,q-1}-(N-1)]^{1/(1-q)} \,.
\end{eqnarray}
For $0 <p<1$ we see that $r_{N,0}=p^N=e^{-N \ln(1/p)}$ if $q=1$, whereas $r_{N,0}\sim \frac{1}{[(1/p)^{1-q}-1]^{1/(1-q)}} \frac{1}{N^{1/(1-q)}}$ $\propto 1/N^{1/(1-q)}$ ($N \to\infty$) for $q<1$. We shall from now on focus on $p=1/2$, and $0 \le q \le 1$. We can check that $q=1$ recovers the triangle on which de Moivre and Laplace based their historic theorem. We can also check that $q=0$ yields $r_{N,0}=1/(N+1)$, i.e., we recover the Leibnitz triangle \cite{Polya}. 

By introducing a conveniently centered and rescaled variable, namely  $x \equiv \frac{n-(N/2)}{N/2}$, we can check that, for increasing $N$, our set of probabilities approaches (see Figs. 4 and 5) a double-branched $q_e$-Gaussian probability $p(x) \propto e_{q_e}^{-\beta\, x^2}$ with
\begin{equation}
q_e=2-\frac{1}{q} \,.
\end{equation}
See Fig. 6.
We verify that $q=1$, hence $q_e=1$, reproduces the de Moivre-Laplace theorem. For $q$ decreasing from unity to zero, $q_e$ decreases from unity to minus infinity. By double-branched we mean that what we get is two branches of $q_e$-Gaussian with slightly different coefficients $\beta$ on the right and left sides of it. This asymmetry comes from the fact that we have introduced the correlations by imposing Leibnitz rule on the ``left" side of the triangle, not, for instance, at its center.

Transformation (8) is a composition of {\it multiplicative duality} ($q \leftrightarrow 1/q$, whose fixed point is $q=1$), and {\it additive duality} ($q \leftrightarrow 2-q$, whose fixed point once again is $q=1$). These two transformations, alone or combined, appear in fact very frequently in the literature of nonextensive statistical mechanics.

\section{3 - Conclusions}

It is well known that, if we have a system constituted by $N$ elements that are (either exactly or nearly exactly) {\it independent} in the probabilistic sense, then (i) the entropy which is extensive is $S_{BG}(N)$, and (ii) the attractor, in the sense of a central limit theorem, is a Gaussian (L\'evy) distribution if the variance of the single distribution that is being convoluted is finite (infinity).    
 
We have shown here that special {\it global correlations} (which are either exactly or asymptotically scale-invariant) among these $N$ elements can make that (i) the value of $q$ for which $S_q(N)$ is extensive differs from unity, and (ii) the attractor, in the sense of a central limit theorem, not only differs from both a Gaussian or a L\'evy distribution, but can even precisely be a $q$-Gaussian distribution (which in fact extremizes $S_q$ under appropriate constraints). 

It remains as an important open question the understanding of what are the precise classes of global correlations that induce one or the other of these anomalies. It is in principle possible that one of these two classes contains the other one, or that none of them fully contains the other one, having nevertheless a nontrivial intersection. Or even --- the simplest of all possibilities --- they could coincide. If they do, is it (asymptotic) scale-invariance a sufficient and necessary condition for both anomalies to emerge simultaneously? The exact answer to this and related questions would deeply enlighten the reason for which $q$-exponentials appear so frequently in many natural and artificial systems. It would of course also qualify the classes of situations for which the concepts emerging in nonextensive statistical mechanics (and in its generalizations and its variations) are applicable.   

\section*{Acknowledgments}
In the expression of the ideas contained in this paper, I have benefitted along many years from interesting discussions with a group of colleagues which includes C. Anteneodo, F. Baldovin, E.P. Borges, E.M.F. Curado, M. Gell-Mann, L.G. Moyano, S.M.D. Queiros and A. Robledo, among many others. The same has happened more recently with  J.D. Farmer, M. Fuentes, J.M. Gomez-Soto, R. Hersh, J. Marsh,  Y. Sato, S. Steinberg and S. Umarov.  Financial support from SI International and AFRL is acknowledged as well.


\end{document}